\documentclass[pof]{revtex4}  

\usepackage{graphicx}  % needed for figures
\usepackage{dcolumn}   % needed for some tables
\usepackage{bm}        % for math
\usepackage{amssymb}   % for math
\usepackage{color}
\usepackage{setspace}
\usepackage{amsmath}
\usepackage{longtable}
\usepackage{subfigure}
\usepackage{ulem}

\begin{document}

\title{Intermittency in 2D soap film turbulence}

\author{R.T. Cerbus}
\email{rtc17@pitt.edu}
\affiliation{Department of Physics and Astronomy, University of Pittsburgh, 3941 O'Hara Street, Pittsburgh PA 15221}
\author{W.I. Goldburg}
\affiliation{Department of Physics and Astronomy, University of Pittsburgh, 3941 O'Hara Street, Pittsburgh PA 15221}

\begin{abstract}

The Reynolds number dependency of intermittency for 2D turbulence is studied in a flowing soap film. The Reynolds number used here is the Taylor microscale Reynolds number $R_{\lambda}$, which ranges from 20 to 800. Strong intermittency is found for both the inverse energy and direct enstrophy cascades as measured by (a) the pdf of velocity differences $P( \delta u(r))$ at inertial scales $r$, (b) the kurtosis of $P(\partial_x u)$, and (c) the scaling of the so-called intermittency exponent  $\mu$, which is zero if intermittency is absent. Measures (b) and (c) are quantitative, while (a) is qualitative. These measurements are in disagreement with some previous results but not all. The velocity derivatives are nongaussian at all $R_{\lambda}$ but show signs of becoming gaussian as $R_{\lambda}$ increases beyond the largest values that could be reached. The kurtosis of $P(\delta u(r))$ at various $r$ indicates that the intermittency is scale dependent. The structure function scaling exponents also deviate strongly from the Kraichnan prediction. For the enstrophy cascade, the intermittancy decreases as a power law in $R_{\lambda}$. This study  suggests the need for a new look at the statistics of 2D turbulence.

\end{abstract}

\maketitle

\section{Introduction}

Three-dimensional turbulence appears in bursts, a phenomenon called ``intermittency''. Said in another way, the flow is characterized by velocity fluctuations of size $r$ that deviate strongly from their mean value. The issue addressed in this experimental study is the presence and strength of intermittency in flows that are very close to being two-dimensional.     

There are several prior studies, both experimental and numerical, that have addressed this issue, but there is little agreement. These experiments support the existence of strong intermittency in two-dimensional (2D) turbulence. In fact they imply that it is even stronger in 2D than in three dimensions (3D). 

Turbulent flow in 2D is decidedly different from its 3D counterpart. In 3D, eddies of size $r$ break up into smaller and smaller ones (say, of size $r/2, r/2^2, r/2^3$...). The total vorticity amplitude is amplified by local velocity gradients, a well-studied effect called ``vortex stretching'' \cite{frisch1995}. On the other hand, in 2D the total squared vorticity $\langle {\bf \omega}^2 \rangle$ (enstrophy) is a constant of the motion in the absence of viscosity, just like the total kinetic energy density $\langle u^2 \rangle$ in 3D. It was R. Kraichnan who first recognized that this new conservation law comes into play with the result that energy is transferred to large scales while enstrophy is handed down to small scales \cite{kraichnan1967a,kraichnan1980,boffetta2012,kellay2002,tabeling2002a}.

Assume that the turbulence is forced at a scale $r = r_f$. In the steady state this forcing results in a mean rate of kinetic dissipation per unit mass $\epsilon = d \langle u^2 \rangle/dt$ and a mean rate of dissipation of the enstrophy, $\beta = d \langle \omega ^2 \rangle/dt$. The Kraichnan analysis gives the result for the longitudinal velocity fluctuations or second-order structure functions
\begin{subequations}
\begin{align}
S_2(r) \equiv \langle \delta u(r)^2 \rangle  \propto \epsilon ^{2/3} r^{2/3} , \,\,\,  r>r_f \\
S_2(r) \equiv \langle \delta u(r)^2 \rangle  \propto  {\beta}^{2/3} r^2 , \,\,\,\, r < r_f
 \label{mlett:2}
\end{align}
\end{subequations}
where $\langle \delta u(r) \rangle \equiv \langle u(x+r) - u(x) \rangle_x$ is the velocity difference on a scale $r$ averaged over $x$ (flow direction). The two cascades do not extend to infinity in both directions. The energy cascade eventually reaches a scale $r = r_{\alpha}$ on the order of the system size where it is cut off by boundary friction or coupling to the 3D environment \cite{boffetta2012}. Likewise, the enstrophy cascade eventually reaches a scale $r = \eta \equiv \nu^{1/2}/\beta^{1/6}$ where viscosity takes over.

A simplified picture of the double cascade process is to view eddies of size $r > r_f$ combining to create the energy cascade. The interstices between these eddies contain most of the shear (and hence large vorticity) but little energy. The result is a cascade of enstrophy to small scales. Recently this picture has come into question \cite{chen2006,chen2003}, but there is full agreement on the strong difference between 3D and 2D turbulence. Both of these cascades are studied in these experiments, although they do not appear simultaneously. This is the case in nearly all 2D experiments and simulations \cite{boffetta2012,kellay2002,tabeling2002a}. Some high resolution simulations and a few experiments are able to achieve the classical dual cascade picture \cite{boffetta2007,boffetta2010,rutgers1998,daniel2000,bruneau2005}.

Intermittency is not a sharply defined effect in 2D or 3D.  The presence of strong, unlikely  velocity fluctuations is manifest in many different types of analysis. Most measures of intermittency are defined in terms of a deviation from the predictions of the scaling theory of Kolmogorov (K41) in 3D \cite{kolmogorov1941} or Kraichnan (Kr67) in 2D \cite{kraichnan1967a}, although intermittency has a larger meaning than this \cite{frisch1995}. Some of those measurements are made in the ``inertial range'' of velocity fluctuations, defined as those eddy sizes where damping effects are negligible. The K41 and Kr67 predictions are for this range of scales. On the other hand, intermittency is reported in measurements of the probability density function (pdf) of velocity gradients $\partial_x u$ or velocity differences $\delta u(r)$. The pdfs are expected to be nearly gaussian if there is no intermittency, with large fluctuations being manifested in the tails. For values of $r$ in the inertial range, the assumptions of K41 and Kr67 also require the shape of the pdfs to be independent of $r$, i.e. self-similar. The statistics of the velocity gradients provide a measure of the ``smoothness" of the velocity field at very small scales outside the inertial range. For 3D energy and 2D enstrophy cascades, these scales lie in the dissipative (viscous) range. Although these measurements of $\partial_x u$ are perhaps not in contradiction to K41 or Kr67, it is still surprising \cite{sreenivasan1997,kraichnan1967b}.

Whereas intermittency is well-established for 3D turbulent flows, there is little agreement about its presence or origin (if present) in 2D turbulence. It is important to consider the two cascades separately. Numerical simulations by Boffetta {\it et al.} and experiments in salt layers by Paret and Tabeling (PT) suggest that the inverse energy cascade is intermittency-free \cite{boffetta2000,paret1998}. An experiment by Jun and Wu (JW) involving a horizontal soap film at a large Reynolds number indicates the opposite \cite{jun2005}. The soap film experiments of Daniel and Rutgers (DR), which exhibit a dual cascade, indicate the presence of intermittency in both \cite{daniel2000}.

The role of 3D effects (in this case air drag) is also not clear. In some simulations it is responsible for intermittent effects in the inertial scales of the enstrophy cascade \cite{boffetta2002,nam2000,tsang2005}. The same linear drag force produces no such effect in numerical simulations of the inverse cascade and is necessary to establish a steady state \cite{boffetta2000,boffetta2012}. Moreover, there is negligible intermittency in salt layer experiments for either the energy or enstrophy cascade despite significant 3D effects from the floor of the container \cite{paret1998,paret1999}. However, a systematic study of the effects of drag on the enstrophy cascade indicate the opposite \cite{boffetta2005}. Other researchers claim that the dissipative scales of the enstrophy cascade are intermittency-free but that coherent structures emerge to produce inertial range intermittency \cite{benzi1986,jimenez2007}. The possible effect of Marangoni stresses, which are present in soap films, is also unknown \cite{chakraborty2011}.

This study focuses on three measures of intermittency. They are (a) the shape of the pdf of velocity differences $\delta u(r)$ at inertial scales $r$ to determine if they are nongaussian and self-similar, (b) the kurtosis of the pdf of $\partial_x u$ and $\delta u(r)$ and (c) measurements of an exponent $\mu$ that characterizes the structure functions $S_n(r)$. Measures (b) and (c) are quantitative, while (a) is qualitative. 

Assuming an appreciable range of inertial scales $r$ where $S_n(r)$ is of algebraic form
\begin{equation}
S_n(r) \equiv \langle | \delta u(r) ^n | \rangle \sim r^{\zeta_n},
\label{zeta_ndefined}
\end{equation}
one defines the intermittency exponent $\mu$ as
\begin{equation}
\mu = \frac{2 \zeta_3 -\zeta_6}{\zeta_3}.
\label{mudefined}
\end{equation}
In the absence of intermittency, $\zeta_n / \zeta_3 = n/3$ for both cascades and so $\mu = 0$. The parameter $\mu$ was first introduced as the fitting parameter for the lognormal model of intermittency and is also the codimension of the dissipation in 3D \cite{frisch1995}. Normalizing by $\zeta_3$ will enable a direct comparison of the degree of intermittency between the cascades. The same normalization choice was made by PT and JW \cite{paret1998,jun2005}. In this work, as in other studies, the average is taken outside the absolute value of the velocity difference. This makes no difference for the even moments but it is not clear why this should be valid for the odd moments \cite{sreenivasan1997}.

An important feature of the intermittency studied here is its strong Reynolds number dependence. Here we use the Taylor microscale Reynolds number $R_{\lambda} \equiv u' \lambda / \nu$,  where $u'$ is the rms velocity and $\lambda = u'/\sqrt{\langle \partial_x u)^2\rangle}$ is the Taylor microscale. For unknown reasons the energy and enstrophy cascade are here observed to occur for low and high $R_{\lambda}$ respectively (see Fig. \ref{derivativeflatness} below). Most previous studies considered only one $R_{\lambda}$ or do not quote it \cite{boffetta2000,boffetta2002,paret1998,nam2000,tsang2005,jun2005,benzi1986}. (A recent simulation claims intermittency in 2D turbulence is $R_{\lambda}$-independent, but the only measures considered are the pdfs of the vorticity and velocity but not their derivatives or structure functions \cite{bracco2010}.) Intermittency measures are $R_{\lambda}$-dependent in 3D, so there is little surprise that the same should be true for 2D \cite{kahalerras1998,sreenivasan1997,jimenez2007}. This being the case, one should take care in comparing results for intermittency in 2D turbulence without reference to $R_{\lambda}$.

\section{Experimental Setup}

\begin{figure}[t!]
%\hspace{-1.5em}
\includegraphics[scale = 0.75]{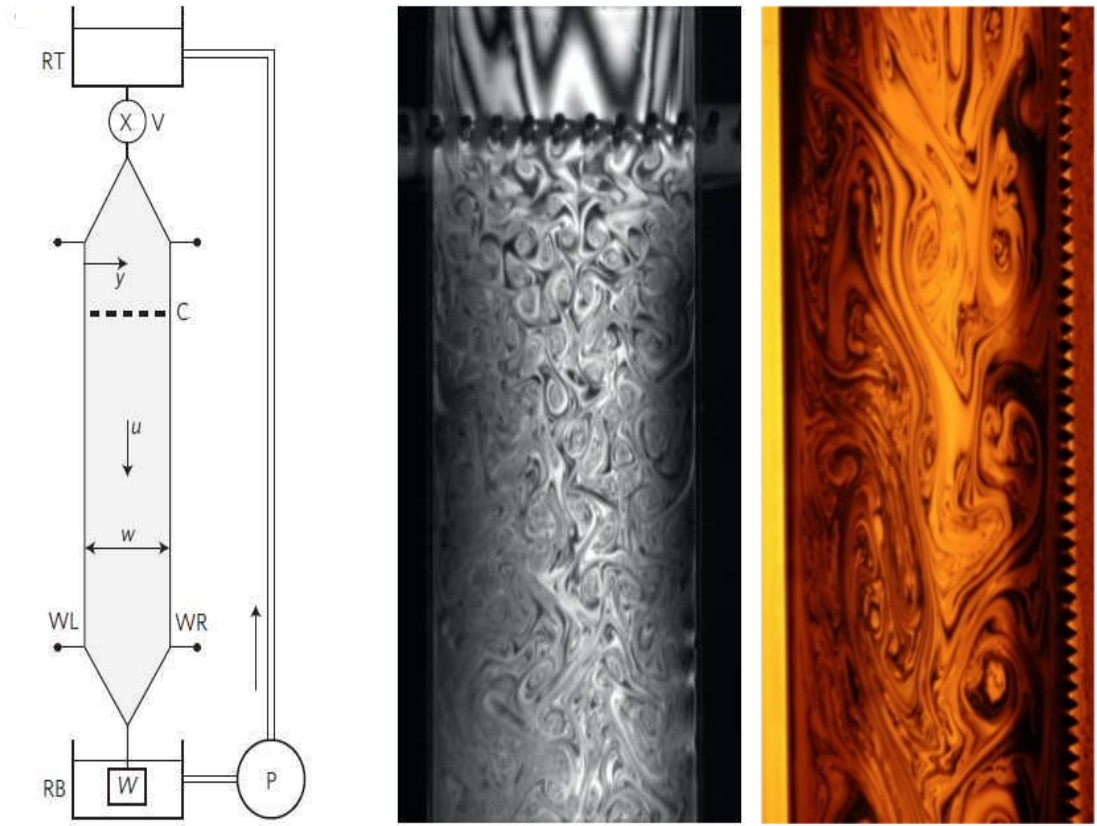}
\caption{Left: Experimental setup showing the reservoirs ($RT$, $RB$), pump ($P$), valve ($V$), comb ($C$), blades ($WL$, $WR$) and weight ($W$). Middle: Fluctuations in film thickness from turbulent velocity fluctuations with smooth walls and a comb. Right: Thickness fluctuations with smooth and rough wall. (Figure adapted from \cite{tran2010}.)}
\label{setup}
\end{figure}

The experimental setup is diagrammed in Fig. \ref{setup}. A soap solution in reservoir $RT$ flows between two blades into reservoir $RB$, where it is pumped back to $RT$ by a small pump $P$, keeping the pressure head constant. The blades have a separation $w$ that is adjustable in the cm range, with a separation of 1-2 cm for most experiments. The blades, which are made of stainless steel, have a thickness of 0.7 mm. The soap solution is 2\% Dawn$^\copyright$ dish-washing detergent in tap water. The valve $V$ regulates the soap solution flux $\Phi$, which is in the range of 0.2-1 ml/s. The film thickness $h$ is roughly 2-20 $\mu$m. The turbulence is generated by a comb usually located 10 cm below the point where thin nylon wires are joined at the top. The spacing of the comb teeth is 2 mm and their thickness is 1 mm. A laser Doppler velocimeter (LDV) permits monitoring the film velocity as function of time at a lateral distance $y$ from a blade. The soap solution contains 4 $\mu$m polystyrene particles that scatter light into the LDV's photodetector. These seed particles are small enough to follow the local velocity fluctuations, their Stokes number being in the range of 0.1 \cite{merzkirch1987}. As discussed below, one of the smooth blades is sometimes replaced by a serrated blade having an indentation depth of 2 mm. In this case the comb is replaced by a single 1 mm rod thrust through the film.

The soap film typically flows with mean vertical velocity (averaged over the film width) of $U \simeq 2$ m/s, with the rms velocity fluctuations $u'$ roughly 0.1 $U$, so the turbulent intensity ${\cal I}  = u'/U \simeq$ 0.1. Replacing one smooth blade with a serrated one, and by varying $\Phi$ and $w$, ${\cal I}$ becomes an adjustable parameter, as does $R_{\lambda}$. The ratio ${\cal I}$ is sufficiently small that one can use the Taylor frozen turbulence hypothesis, enabling translation of temporal shifts into spacial displacements \cite{belmonte2000}. In this work only longitudinal velocities are considered. However, a few sets of transverse data, which are of lesser quality, show similar behavior. This, along with similar rms values, suggests that the flow is reasonably isotropic.

The mean data rate is typically 5000 Hz, making the smallest scales of the inertial range easily resolvable. The data are interpolated for equal spacing. To ensure insensitivity to the interpolation method, several are used. Linear and nearest neighbor schemes (similar to sample and hold \cite{ramond2000}) are used in conjunction with the usual Taylor hypothesis: $x = tU$. The third method introduced by Kahalerras {\it et al.} seeks to correct for biases in the sampling stemming from the fluctuations in velocity \cite{kahalerras1998}. No significant difference is found between these different interpolation schemes.

Using the above methods, one extracts the structure functions  $S_n(r)$ and the energy spectrum $E(k)$ at desired points ($x,y$) in the soap film \cite{pope2000}.  The measurements are usually made 10 cm below the comb, but the distance is not found to be important. Taking the Fourier transform of $u$ yields the one-dimensional energy spectrum $E_1 (k)$ \cite{pope2000}. If the turbulence is isotropic, $E(k) = 10 E_1(k)$.

Energy spectra are measured using either two smooth walls or one smooth and one rough wall. The enstrophy cascade is observed when smooth walls are used, and the inverse energy cascade is usually present when one smooth wall is replaced by a serrated one. This presumably changes decaying turbulence to forced turbulence and is the method recently used to study friction in the energy cascade \cite{kellay2012}. 

\begin{figure}[h!]
%\hspace{-1.5em}
\includegraphics[scale = 0.5]{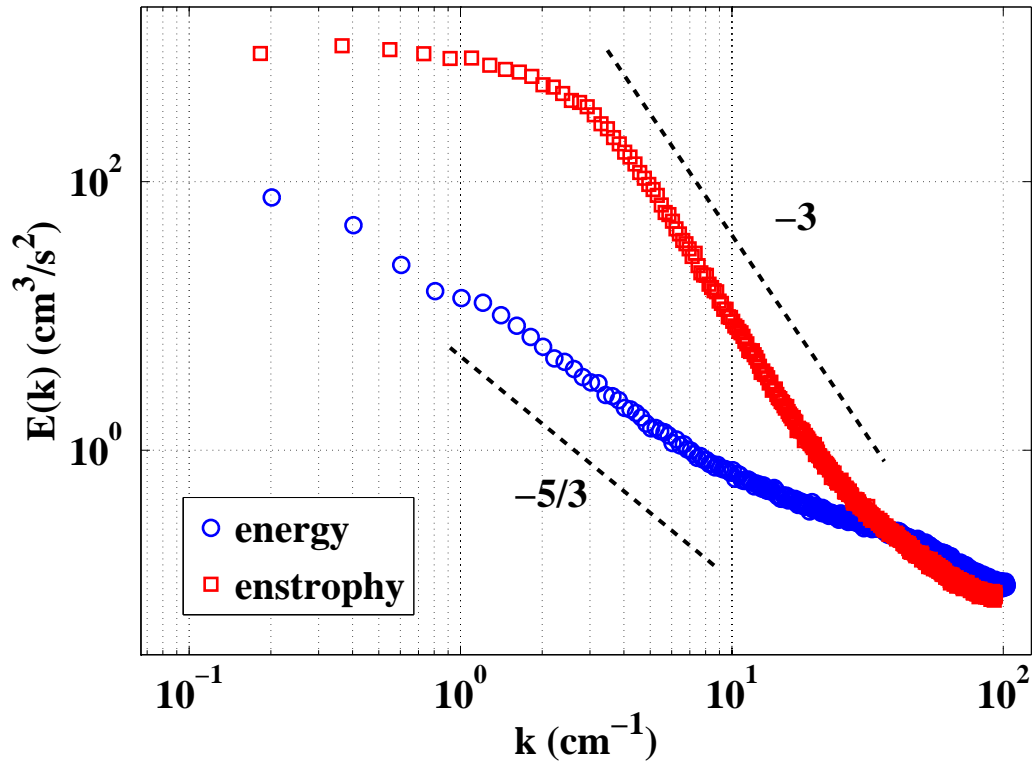}
\caption{ Energy spectra $E(k)$ in cm$^3$/s$^2$ measured at the centerline. The upper curve ($\square$) is the enstrophy cascade and the lower curve ($\bigcirc$) is the energy cascade data.  While they are only guides to the eye, the straight lines are the expected slopes for the energy and enstrophy cascades (Kr67 or dimensional reasoning).  The spectra are normalized such that $\int_{0}^{\infty} dk E(k)\ = \frac{1}{2} \langle {\bf u}^2\rangle $.}
\label{spectra}
\end{figure}

The energy and enstrophy spectra are seen in Fig. \ref{spectra}. Though the slopes $\gamma$ in the scaling relation $E(k) \sim k^{-\gamma}$ are only approximately equal to the canonical values $\gamma$ = 5/3 (energy) and $\gamma$ = 3 (enstrophy), they will be referred to as such. The reason for  the departure from 5/3 and 3 is unclear,  but it may be related to the intermittency studied here. Additional evidence that the $\gamma = 5/3$ spectra really correspond to the inverse energy cascade comes from the third order structure function $S_3(r)$ (where the absolute value has not been taken). See Appendix A for the details.

\section{Results}

\subsection{Probability Density Functions}

\begin{figure}
\hfill
\subfigure[$$ energy cascade, $R_{\lambda}$ = 50]{\includegraphics[scale = 0.3]{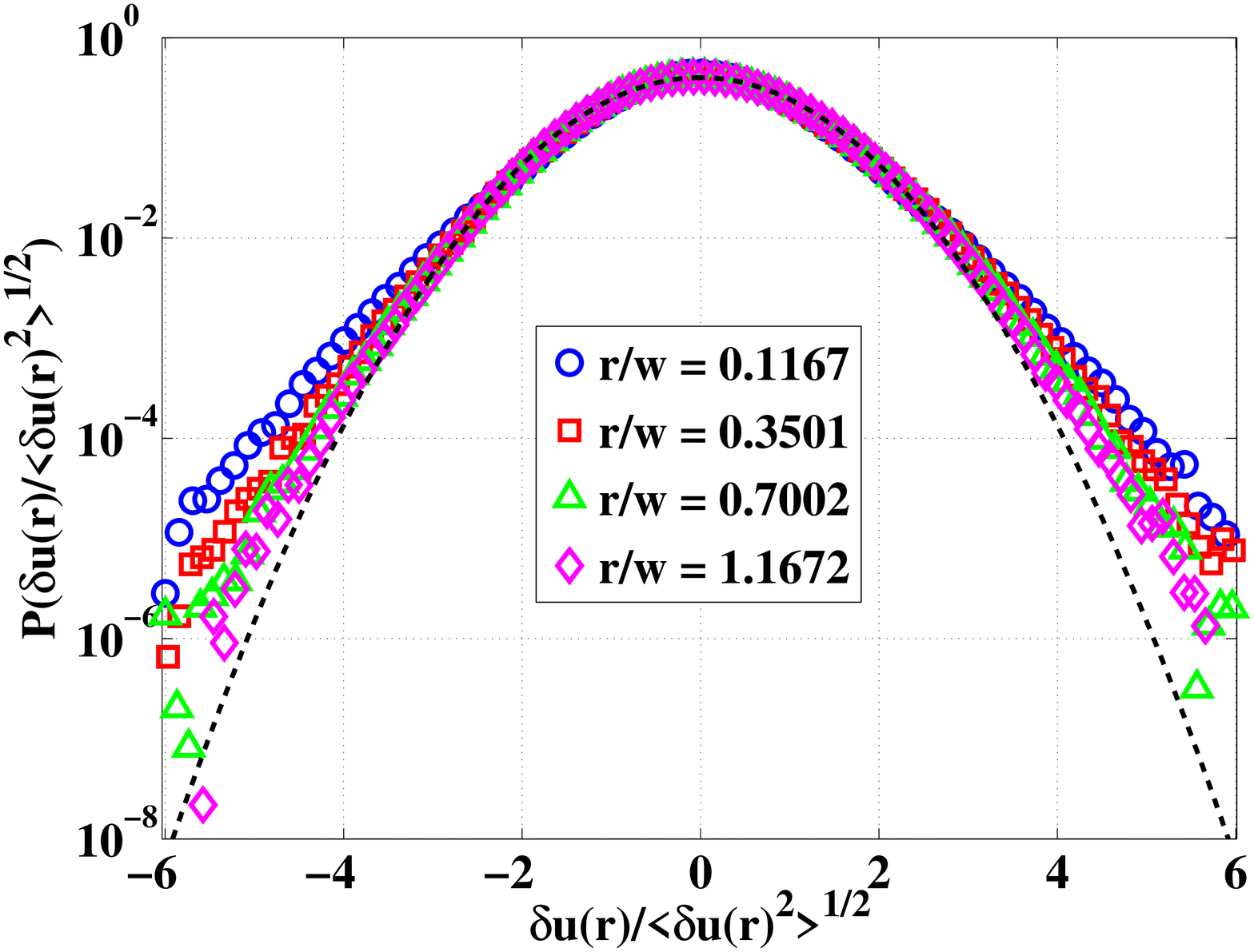}}
\hfill
\subfigure[$$ enstrophy cascade, $R_{\lambda}$ = 610]{\includegraphics[scale = 0.3]{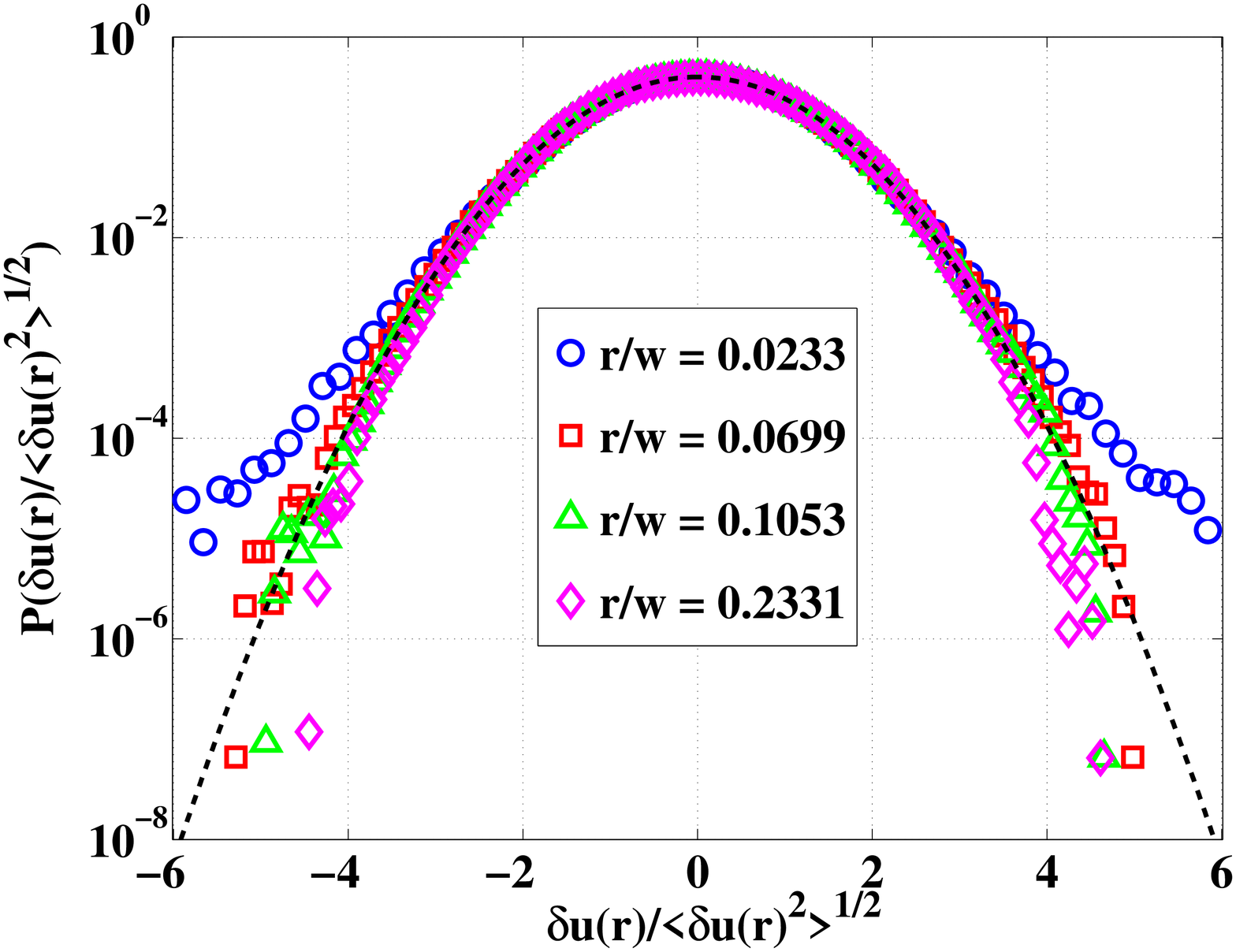}}
\hfill
\caption{Probability density functions (pdfs) of longitudinal velocity differences $\delta u(r)$ measured for the energy cascade (a) at $R_{\lambda}$ = 50, and for the enstrophy cascade (b) at $R_{\lambda}$ = 610. The first three values of $r$ here are in the inertial range as determined by the power law scaling region of the structure functions. The dashed line is a gaussian function with zero mean and a standard deviation of unity. The mean and variance of the velocity data have been normalized so that if they are gaussian, they will lie on top of this curve. None of these pdfs can be truly gaussian as their third moments cannot vanish \cite{belmonte1999}, but the energy cascade data deviate much more than the enstrophy cascade data. The pdfs at different $r$ do not have the same shape, indicating a lack of self-similarity.}
\label{pdf}
\end{figure}

Figure \ref{pdf} shows the normalized longitudinal velocity difference probability $P(\delta u(r) / \sqrt{\langle \delta u(r)^2  \rangle})$ as a function of the dimensionless velocity difference $\delta u(r)/\sqrt{\langle \delta u(r)^2\rangle}$ for several values of $r$ for both cascades. Data for the energy cascade (a) and the enstrophy cascade (b) are shown. Both pdfs must be skewed and hence not perfectly gaussian to assure that the energy or enstrophy flux is nonzero \cite{belmonte1999}. It is apparent from this qualitative measure of intermittency that the energy cascade pdf has wider flanks than the enstrophy cascade pdf and is notably nongaussian. On the other hand, the enstrophy cascade pdf is well-fitted by a gaussian function (dashed line). The relative difference between the $R_{\lambda}$-values is significant. Some enstrophy data with smaller $R_{\lambda}$ do not have a perfect gaussian distribution.

The pdfs at different $r$ have measurably different shapes. This indicates a lack of self-similarity, since the statistics of velocity fluctuations should be the same at all scales $r$ in the inertial range. The absence of self-similarity provides additional evidence that the flows are intermittent. The classical Kr67 theory of 2D turbulence (as well as the K41 theory of 3D turbulence) assumes self-similarity in the inertial range \cite{kraichnan1967a,frisch1995}.

Although these measures of intermittency are qualitative, they signal a disagreement with the most recent and prevalent view of 2D intermittency \cite{boffetta2012}. That is, it is held that the energy cascade should be non-intermittent while the enstrophy cascade is necessarily intermittent (for a variety of reasons). The quantitative measures will be investigated next.
  
\subsection{Flatness}

\subsubsection{Velocity Derivative Flatness}

The flatness is now considered as the first quantitative measure of intermittency. The velocity derivative $\partial_x u$ in turbulence plays a special role compared to inertial-scale velocity differences $\delta u(r)$. Experiments in 3D show that the mean viscous energy dissipation rate per unit mass  $\epsilon_{\nu}$, which is given by
\begin{equation}
 \epsilon_{\nu} = \nu \langle { |\nabla {\bf u}|^2} \rangle,
 \label{meandissipationrate}
\end{equation}
approaches a constant as $R_{\lambda} \rightarrow \infty$ \cite{frisch1995}. Imagine increasing $R_{\lambda}$ to infinity by taking the limit $\nu \rightarrow 0$.  It is then necessary for the velocity gradients to diverge in order to retain the constancy of $\epsilon_{\nu}$. This can occur if  the pdf of velocity gradients  develop increasingly wider skirts.  Define the derivative flatness as
\begin{equation}
F_{\eta} \equiv  \frac{ \langle (\partial_x u) ^4 \rangle }{ \langle (\partial_x u)^2 \rangle ^2}.
\end{equation}
A broader pdf for $\partial_x u$ means a larger $F_{\eta}$. Keep in mind that $F_{\eta}$ is a measure of small scale intermittency. A question posed by 3D experiments, and by these 2D experiments as well, concerns the dependence of $F_{\eta}$ on $R_{\lambda}$. In 3D it is found that $F_{\eta}$ increases monotonically with $R_{\lambda}$ in a self-similar fashion \cite{sreenivasan1997}, eventually flattening out \cite{tabeling2002b}. In practice the velocity derivative must be approximated by a velocity difference on a very small scale. It should be smaller than the dissipative scale $\eta$ for the enstrophy cascade. Although the smallest scale for the energy cascade inertial range $r_f$ is not well-defined in these experiments, the smallest scales probed here are significantly smaller than the smallest scale (highest wavenumber $k$) in the power law region of $E(k)$. Moreover, $F_{\eta}$ may have no meaning for the energy cascade, since it probes intermittency in scales much smaller than  inertial range scales.  

The mean data rate to which the measurements are interpolated is high enough that velocity derivatives are calculated at a scale smaller than the inertial range scales for each case considered. The velocity derivative is estimated using the central difference method ($\partial_x u \simeq \frac{u_{i+1} - u_{i-1}}{x_{i+1} - x_{i-1}}$). (The simpler adjacent point method gives slightly higher values of $F_{\eta}$.)

\begin{figure}[h!]
%\hspace{-1.5em}
\includegraphics[scale = 0.5]{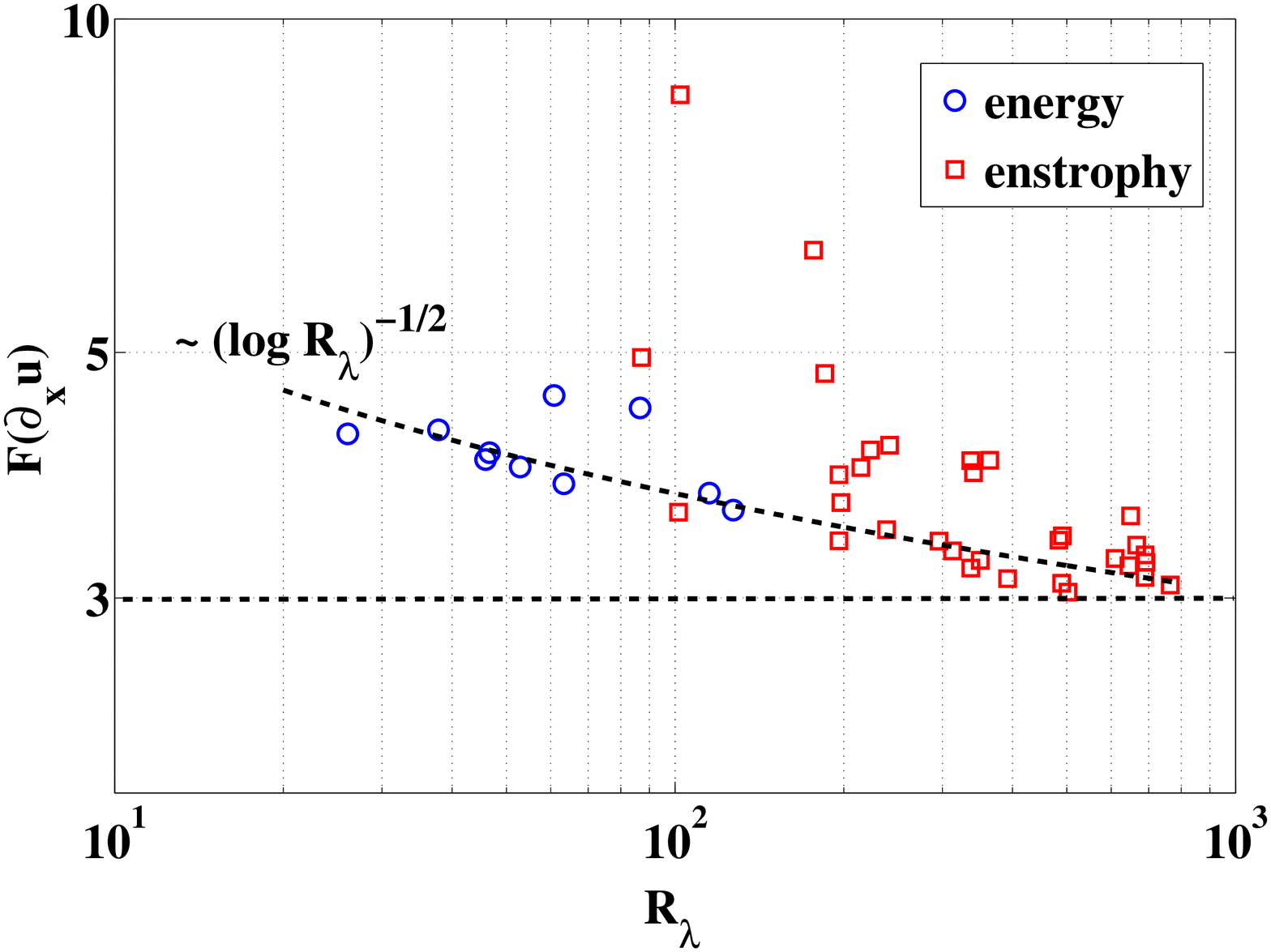}
\caption{Flatness of the velocity derivative $F_{\eta}$ vs. $R_{\lambda}$ with $\partial_x u$ estimated using the central difference method. The energy cascade data ($\square$) and the enstrophy cascade data ($\triangle$) fall into two sections with distinct $R_{\lambda}$. A curved dashed line is shown ($f(R_{\lambda}) = 8(\log R_{\lambda})^{-1/2}$) which suggests a (sub-) logarithmic approach to zero.}
\label{derivativeflatness}
\end{figure}

Figure \ref{derivativeflatness} shows $F_{\eta}$ vs. $\log R_{\lambda}$. The data span a wide range of $R_{\lambda}$ and include both energy and enstrophy data. There is a perceptible downward trend (note dashed line), but the scatter is indeed large, especially for the enstrophy data. In the energy range, $F_{\eta}$ is larger than the gaussian value of 3 and is comparable to 3D values at the same $R_{\lambda}$ \cite{tabeling2002b}.

One might extrapolate this downward trend in $F_{\eta}(R_{\lambda})$ to infer that as $R_{\lambda} \rightarrow \infty$, 2D turbulence has no small-scale intermittency. There seem to be no previous studies of small-scale intermittency for 2D turbulence, but in 3D both experiments and simulations show that $F_{\eta}$ is an increasing function of $R_{\lambda}$ before possibly leveling off at a value larger than 3 \cite{tabeling2002b}. The $R_{\lambda}$-independence of $\epsilon_{\nu}$ is the likely explanation of this increase of $F_{\eta}$ in 3D turbulence, as discussed previously. In 2D turbulence this dissipative anomaly is absent: $\epsilon_\nu \rightarrow 0$ as $R_{\lambda} \rightarrow \infty$ (or as $\nu \rightarrow 0$) \cite{boffetta2012}. Hence, $F_{\eta}$ need not increase as $R_{\lambda}$ increases. Some claim that $\beta$ asymptotes to zero slower than logarithmically: $\beta \rightarrow 0$ as $(\log R_{\lambda})^{-1/2}$, which may be related to the apparently (sub-) logarithmic vanishing of $F_{\eta}$ with $R_{\lambda}$ (see Fig. \ref{derivativeflatness}) \cite{tran2006,dritschel2007}.

\subsubsection{Inertial Range Flatness}

\begin{figure}[t!]
%\hspace{-1.5em}
\includegraphics[scale = 0.50]{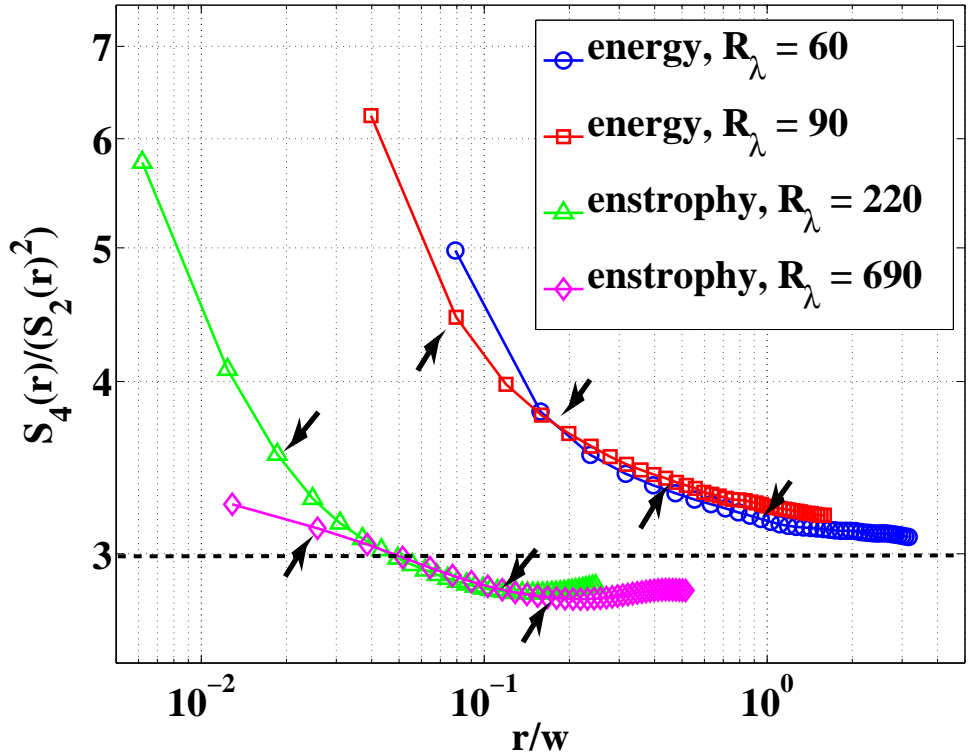}
\caption{Flatness $F(r)$ of $P(\delta u(r))$ depending on the scale $r$ (normalized by channel width $w$). The two upper curves are for energy cascade data ($\bigcirc$: $R_{\lambda}$ = 60, $\square$: $R_{\lambda}$ = 90) and the two bottom curves are for enstrophy cascade data ($\triangle$: $R_{\lambda}$ = 490, $\diamond$: $R_{\lambda}$ = 690). The arrows denote the beginning and end of the inertial range for each curve, as determined by the power law region of the structure functions. All of the curves start out above the gaussian value of 3, but the energy cascade data remain above and seem to asymptote to 3 as $r$ increases. The enstrophy cascade curves cross guassianity in the inertial range and then asymptote to a value smaller than 3.}
\label{flatness_sfs}
\end{figure}

Now consider the flatness of $P(\delta u(r))$ for $r$ at any scale.
\begin{equation}
F(r) \equiv \frac{\langle \delta u(r) ^4 \rangle}{\langle \delta u(r)^2 \rangle ^2} = \frac{S_4(r)}{S_2(r)^2}.
\end{equation}
This quantity allows one to measure the deviations from gaussianity, specifically the strength of rare fluctuations, at various scales $r$, whereas $F_{\eta}$ focuses on the small scales only. Figure \ref{flatness_sfs} shows $F(r)$ for several $R_\lambda$ for both the energy and enstrophy cascades. The values of $r$ span a wide range of scales that include the inertial range (indicated by arrows). The curves with low $R_{\lambda}$ show significant nongaussianity.

The two cascades show rather different behaviors as $r$ increases. The energy cascade data appear to approach 3, indicating that gaussianity is restored at large scales just as in 3D \cite{jimenez2007}. The enstrophy cascade data reach a value of 3 in the inertial range and then continue to decrease to slightly smaller values. For some of the lower $R_{\lambda}$ enstrophy data, there are small bumps at large $r$ outside the inertial range. Since $F(r)$ is larger for the energy cascade than the enstrophy cascade, even in the inertial range, this means that the large scale intermittency is not only present but also stronger in the energy cascade. Moreover, given the direction of the cascades, one may infer that the eddies become intermittency-free as they combine for the energy cascade but become more intermittent as they break up in the enstrophy cascade.

\subsection{Intermittency Exponent}
  
One of the many measures of intermittency involves the structure functions $S _n(r)$.  In the absence of intermittency, $S_n(r) \sim r^{\zeta_n}$, with $\zeta_n = n/3$ for the energy cascade (in 3D or 2D), according to K41 and Kr67 \cite{kolmogorov1941,frisch1995,kraichnan1967a}.  For the 2D enstrophy cascade, $\zeta_n = n$ \cite{kraichnan1980}. Subsequent 3D experiments have shown that $\zeta_n/n$  is not a constant but decreases with increasing $n$. This is one identifier of intermittency \cite{frisch1995}. 

Just as with the velocity derivatives and indeed any measure of intermittency, there are sometimes delicate and controversial issues involved in estimations. The issue of estimating scaling exponents will be discussed below, but before this can be done the reliability of the higher order structure functions should be addressed. As in the experiment of JW \cite{jun2005}, the method advanced by Dudok de Wit is used to indicate the highest reliable order \cite{wit2004}. This method has the advantage of being less subjective than other methods. See Appendix B for a more complete description. Also discussed in Appendix A is an additional technique for determining the reliability of the data. Using Dudok de Wit's method, most data were reliable up to order 12 or 13.
 
The soap film experiments described here add evidence that intermittency is also present in 2D. The strongest prior evidence for its presence in 2D comes from a study of forced steady-state turbulence at $R_{\lambda}$ comparable to that used here. In the experiments of JW \cite{jun2005}, the 2D system is an electrically conducting soap film driven by an array of small magnets placed below it.

\begin{figure}[t!]
%\hspace{-1.5em}
\includegraphics[scale = 0.50]{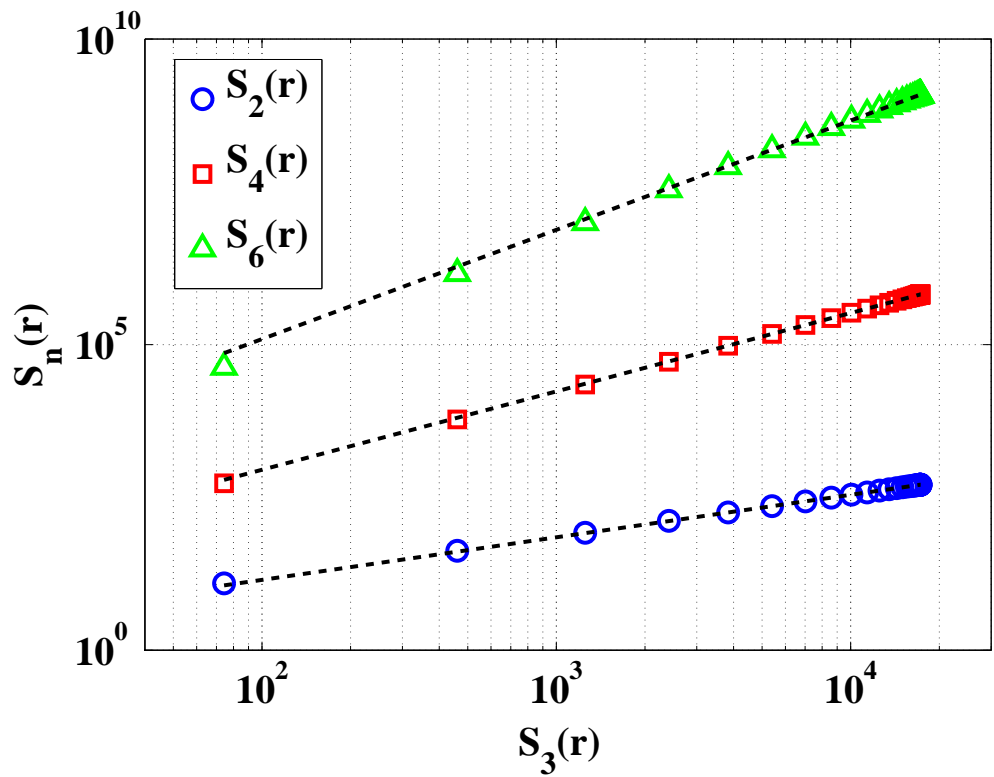}
\caption{Three measured structure functions whose scaling exponents are determined using extended self-similarity (ESS). That is, $S_n(r)$ is plotted vs. $S_3(r)$. From the bottom to the top, the lines are $S_2(r)$, $S_4(r)$ and $S_6(r)$. Here $R_{\lambda} = 490$. The data exhibit the enstrophy cascade.}  
\label{ESS246}
\end{figure}

They evaluated the moments $S_n(r)$ using extended self-similarity (ESS) and several other methods as a check \cite{benzi1993}. With ESS, one plots $S_n(r)$ as a function of $S_3(r)$ rather than $r$ itself. With this approach the range of self-similarity extends over a broad range of the independent variable $S_3(r)$, as compared with a plot of $S_n(r)$ vs. $r$ itself. The same has been done in this study. Figure \ref{ESS246} shows an example with $S_2(r)$, $S_4(r)$, and  $S_6(r)$ vs. $S_3(r)$ for enstrophy cascade data with $R_\lambda$ = 490. The dashed lines in the figure are least square fits to the data.

\begin{figure}[t!]
%\hspace{-1.5em}
\includegraphics[scale = 0.50]{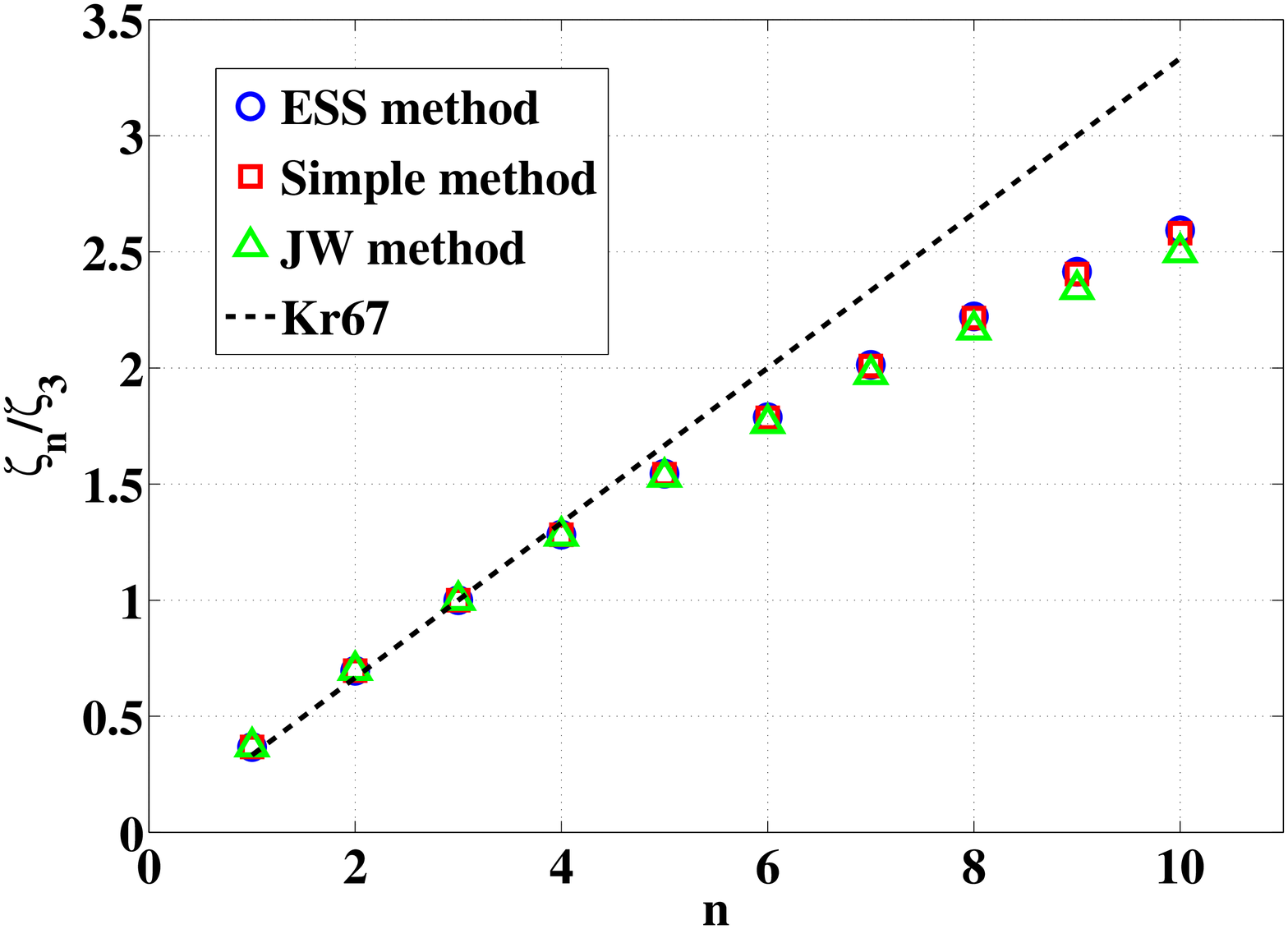}
\caption{Normalized scaling exponents of the $n$th-order structure functions out to $n$ = 10 for the enstrophy cascade with $R_{\lambda} = 490$. The data set denoted by squares ($\square$) is extracted from measurements spanning a decade in $r$. The open circles ($\bigcirc$) denote slopes deduced using extended self-similarity \cite{benzi1993}. The triangles ($\triangle$) denote measurements obtained using the method from JW \cite{jun2005}. All methods agree very well with each other.}
\label{zeta_n}
\end{figure}

Figure \ref{zeta_n} provides a comparison of the various methods for estimating $\zeta_n$. One sees here the normalized structure function exponents $\zeta_n/\zeta_3$ for values of $n$ in the interval 1 through 10.  The open squares ($\square$) designate direct measurements of $\zeta_n/\zeta_3$ from the enstrophy cascade data ($R_{\lambda} = 490$) without invoking ESS. This method is labeled ``Simple". The open circles ($\bigcirc$) are obtained using ESS, as discussed above. The open triangles ($\triangle$) are calculated as $\langle \frac{d \log S_n(r)}{d \log r} / \frac{d \log S_3(r)}{d \log r} \rangle$ just as in JW \cite{jun2005}. The straight dashed line is the Kr67 prediction and has a slope of $\zeta_n/\zeta_3$ = $n/3$ (no intermittency).  It is apparent that the various methods give similar results with any significant differences occurring for larger $n$. The main results for $\mu$ presented here were obtained using the ESS method.

\begin{figure}[t!]
\includegraphics[scale = 0.50]{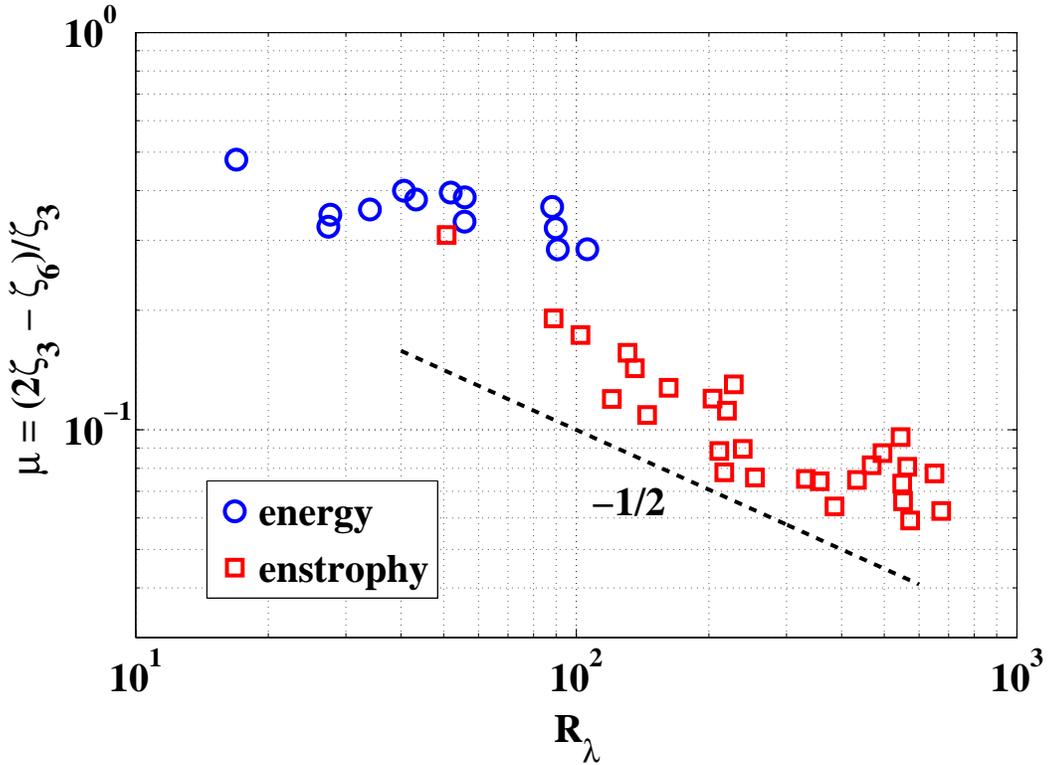}
\caption{Intermittency exponent $\mu$ vs. $R_{\lambda}$ for energy cascade ($\bigcirc$) and the enstrophy cascade ($\bigcirc$). The value of $\mu$ for the energy cascade is roughly constant while $\mu$ for enstrophy cascade appears to be a decreasing function of $R_{\lambda}$.}
\label{mu}
\end{figure}

In the present study $\mu$ is measured as a function of $R_{\lambda}$, which was not an adjustable parameter in prior measurements (JW's experiment varied $R_{\lambda}$ but focused on only one value \cite{jun2005}).  The results in Fig. \ref{mu} show that the dependence of $\mu$ on $R_{\lambda}$ cannot be ignored. The observations are divided into two sets.  The open circles ($\bigcirc$) denote measurements made for the energy cascade, where  $10 < R_{\lambda} < 110$. The open squares ($\square$) denote measurements made for the enstrophy cascade, where $50 < R_{\lambda} < 700$.  For the energy cascade, the maximum value of  $\mu$ is $\simeq$ 0.5, which is more than twice its value in 3D flows \cite{anselmet1984,sreenivasan1997}. It is notable that $\mu$ depends rather strongly on $R_{\lambda}$, at least for the enstrophy cascade. Just as with $F_{\eta}$, this decrease of $\mu$ with $R_{\lambda}$ suggests that as $R_{\lambda} \rightarrow \infty$, intermittency vanishes for 2D turbulence.

Although there is significant $R_{\lambda}$-dependence for the intermittency, the mean values and variances will roughly indicate the strength for each cascade (energy and enstrophy). Table \ref{tab:tabulation}, which contains the statistics for $F_{\eta}$ and $\mu$, shows that they are statistically reliable.

\begin{table}
\begin{tabular}{ | c | c | c | c | c | }
\hline

 & $\langle F_{\eta} \rangle $ & $ \sigma_{F_{\eta}}$  & $\langle \mu \rangle $ & $ \sigma_{\mu} $ \\
\hline
energy & 4.06 & 0.3 & 0.36 & 0.05 \\
\hline
enstrophy &  3.80 & 1 & 0.11 & 0.05 \\
\hline
\end{tabular}
\caption{Average ($\langle \cdot \rangle$) and standard deviation values ($\sigma$) of intermittency measures for energy and enstrophy data. All mean values show significant deviations from the non-intermittent standard. The enstrophy cascade standard deviations are large due to the $R_{\lambda}$-dependence. Recall that for 3D, $\mu \simeq 0.2$ \cite{frisch1995}.}
\label{tab:tabulation}
\end{table}

The values of $\mu$ measured here are significantly larger than that obtained by PT for the energy cascade \cite{paret1998}. For their magnetically driven salt layer experiment, they found (also using ESS) that $\mu \simeq 0.02$. The value of $\mu$ found by JW was $\mu \simeq 0.11$, which is comparable with the enstrophy cascade values measured here, although JW focused on the energy cascade \cite{jun2005}. The JW experiment quotes $R_{\lambda} \simeq 170$, which is significantly higher than the maximum value obtained in the present energy cascade experiments. This may explain the discrepancy in the measured values of $\mu$. The experiments of DR do not quote $\mu$, but there is a stronger deviation in the exponents in the enstrophy cascade range than the energy cascade \cite{daniel2000}. Their work has the novelty  of looking at a dual cascade, which may be the reason for the disagreement with the present measurements. The results for $\mu$ for the enstrophy cascade with drag obtained by numerical simulations seem comparable to the energy cascade result here (see Table \ref{tab:tabulation}). As mentioned earlier, it may not be meaningful to compare the intermittency between experiments when $R_{\lambda}$ is different. For most previous studies, no $R_{\lambda}$ (or any Reynolds number) is quoted \cite{boffetta2000,boffetta2002,tsang2005,paret1998,paret1999}.

The argument that air drag is responsible for the intermittency in the enstrophy cascade does not deal with its $R_{\lambda}$-dependence. A term linearly proportional to the velocity (or vorticity) is usually added to the 2D Navier-Stokes equation to account for 3D drag effects: 
\begin{equation}
\frac{\partial {\bf u}}{\partial t} + ({\bf \nabla} \cdot {\bf u} ) {\bf u} = - \frac{1}{\rho} {\bf \nabla} p + \nu \nabla^2 {\bf u} - \alpha {\bf u} + {\bf f}
\end{equation}
where ${\bf f}$ is some external forcing and $\alpha$ is the drag coefficient \cite{nam2000,tsang2005,boffetta2000,boffetta2002,rivera2000}. The value of $\alpha$ extracted from a horizontal soap film experiment is $\alpha \simeq 0.7 \pm 0.3$ Hz \cite{rivera2000}, while estimates based on a boundary layer approximation suggest $\alpha \simeq 0.1$ Hz \cite{boffetta2012}. The corrections to the enstrophy cascade scaling exponents are on the order of $\alpha' = \alpha/\omega'$, where $\omega'$ is the rms vorticity \cite{boffetta2005,boffetta2002,tsang2005,nam2000}. Namely, the exponents are modified to become $\zeta_n \simeq n(1+\alpha')$. This correction is estimated to be on the order of $10^{-3}$ for the system used here. This value of $\alpha'$ is in good agreement with that calculated in \cite{boffetta2012} for falling soap films. It is clearly too small to be detectable. Experiments with soap films in partial vacuums, where the air drag has been considerably reduced, indicate that the air drag has little effect on the spectral exponent \cite{martin1998}. The effects of air drag on the inverse energy cascade have not previously been considered since such 3D effects are necessary to maintain the steady state. With regard to coherent structures, their presence is always accompanied by energy spectra $E(k)$ that are much steeper than is observed in this work ($\gamma \ge 5$), so they cannot be the root of the intermittency here \cite{benzi1986}.

\section{Fractal dimension of turbulence and intermittency}

In an attempt to understand why intermittency decreases with increasing
$R_{\lambda}$, the $\beta$ model is invoked ($\beta \neq \omega^2$). The $\beta$ model is based on a cartoonish description of the cascade where ``mother" eddies do not transfer all of their energy or enstrophy to their ``daughter" eddies. According to this model, turbulent fluctuations are not space-filling but rather occupy a spatial dimension $D$ (fractal) that is less than the embedding dimension $d$ (in this case 2) \cite{frisch1995}. Consider the structure functions $S_n(r) \propto r^{\zeta_n}$. The exponents for the energy and enstrophy cascades respectively are
\begin{subequations}
\begin{align}
\zeta_n = \frac{n}{3} + (2-D)(1-\frac{n}{3}) \\
\zeta_n = n + (2-D)(1-\frac{n}{3}).
 \label{exponent}
\end{align}
\end{subequations}
The energy spectrum is similarly altered for both the energy and enstrophy cascades respectively
\begin{subequations}
\begin{align}
E(k) \sim k^{-\frac{5}{3} - \frac{2-D}{3}} \\
E(k) \sim k^{-3 - \frac{2-D}{3}}.
 \label{spectrumfix}
\end{align}
\end{subequations}
In terms of the intermittency exponent, one has $\mu_{\epsilon} = 2-D$ for the energy cascade and $\mu_{\beta} = \frac{1}{3}(2-D)$ for the enstrophy cascade. This establishes a link between the intermittency measured here and the fractal dimension of the turbulence. Taking $D$ equal to the embedding dimension $d = 2$, gives $\mu_{\epsilon} = \mu_{\beta}$ = 0. Using the typical value of $\mu_{\beta} \simeq 0.1$ for the enstrophy cascade or $\mu_{\epsilon} \simeq 0.3$ for the energy cascade, one finds that $D \simeq 1.7$. On the other hand it is well-established that for the enstrophy cascade in soap films, $E(k) \propto k^{-3.3}$ \cite{boffetta2012,kellay2002}, which suggests that $D \simeq 1.1$ and $\mu \simeq 0.6$. These values do not seem realistic and point to the failure of the $\beta$ model to reproduce the numerical values of the intermittency, although the picture of a fractal turbulence is still attractive.

So far there are no measurements establishing that $D$ is a function $R_{\lambda}$, but some of the enstrophy-range measurements discussed above suggest that $D$ approaches 2 with increasing $R_{\lambda}$ (see Figs. \ref{derivativeflatness}, \ref{flatness_sfs} and \ref{mu}). This is equivalent to the enstrophy cascade becoming more space-filling as $R_{\lambda} \rightarrow \infty$. This should be compared with the 3D energy cascade, where it appears that $D$ saturates to a value less than $d = 3$ as $R_{\lambda} \rightarrow \infty$ \cite{frisch1995}.

Using Eq. \ref{exponent}, the flatness of the structure functions goes as $F(r) \propto r^{D-2}$, which indicates that this quantity blows up as $r \rightarrow 0$ if intermittency is present. This is in accord with the data shown in Fig. \ref{flatness_sfs}. However, as in 3D one expects that a finite value is reached as the limit $r \rightarrow 0$, $F(r) \rightarrow F_{\eta}$. With the assumption that the turbulence becomes space-filling as $R_{\lambda} \rightarrow 0$, a multi-fractal formalism then suggests that $F_{\eta} \rightarrow$ constant (see section 8.5 in \cite{frisch1995}), which is in reasonable agreement with the results in Fig. \ref{derivativeflatness}.

\section{Summary} 

Presented here is a study of intermittency in a flowing soap film. In accord with some prior studies and in disagreement with others, the observed intermittency is as strong in 2D as in 3D. The measurements, which span more than two decades in $R_{\lambda}$, explore both the energy and enstrophy cascades. The intermittency is strongest in the energy cascade and is nearly $R_{\lambda}$-independent. While there is intermittency in the enstrophy cascade, it decreases with increasing $R_{\lambda}$, suggesting that it may vanish in the limit $R_{\lambda} \rightarrow \infty$.

The origin of the intermittency is not clear. There have been several explanations proposed in the literature for the direct enstrophy cascade, but they do not account for its strength or deal with its $R_{\lambda}$-dependence. There also seems to be no fundamental reason to expect the inverse energy cascade to be intermittency-free. Some of the difficulties in coming up with a unified picture of intermittency in 2D may also be due to the failure of the Kr67 picture. The present results suggest the need for taking a fresh look at the statistics of 2D turbulence, as some are already doing \cite{chen2006,chen2003,tran2006,dritschel2007}.
      
\section{Acknowledgments}

The authors are grateful to G. Boffetta for clarifying and explaining the physics of the linear drag. The authors also benefitted from discussions with C. C. Liu and P. Chakraborty. This work is supported by NSF Grant No. 1044105 and by the Okinawa Institute of Science and Technology (OIST). R.T.C. is supported by a Mellon Fellowship through the University of Pittsburgh.

\section{Appendix A: Third-order Structure Function}

\begin{figure}[!h]
\centering
\includegraphics[width=0.6\textwidth]{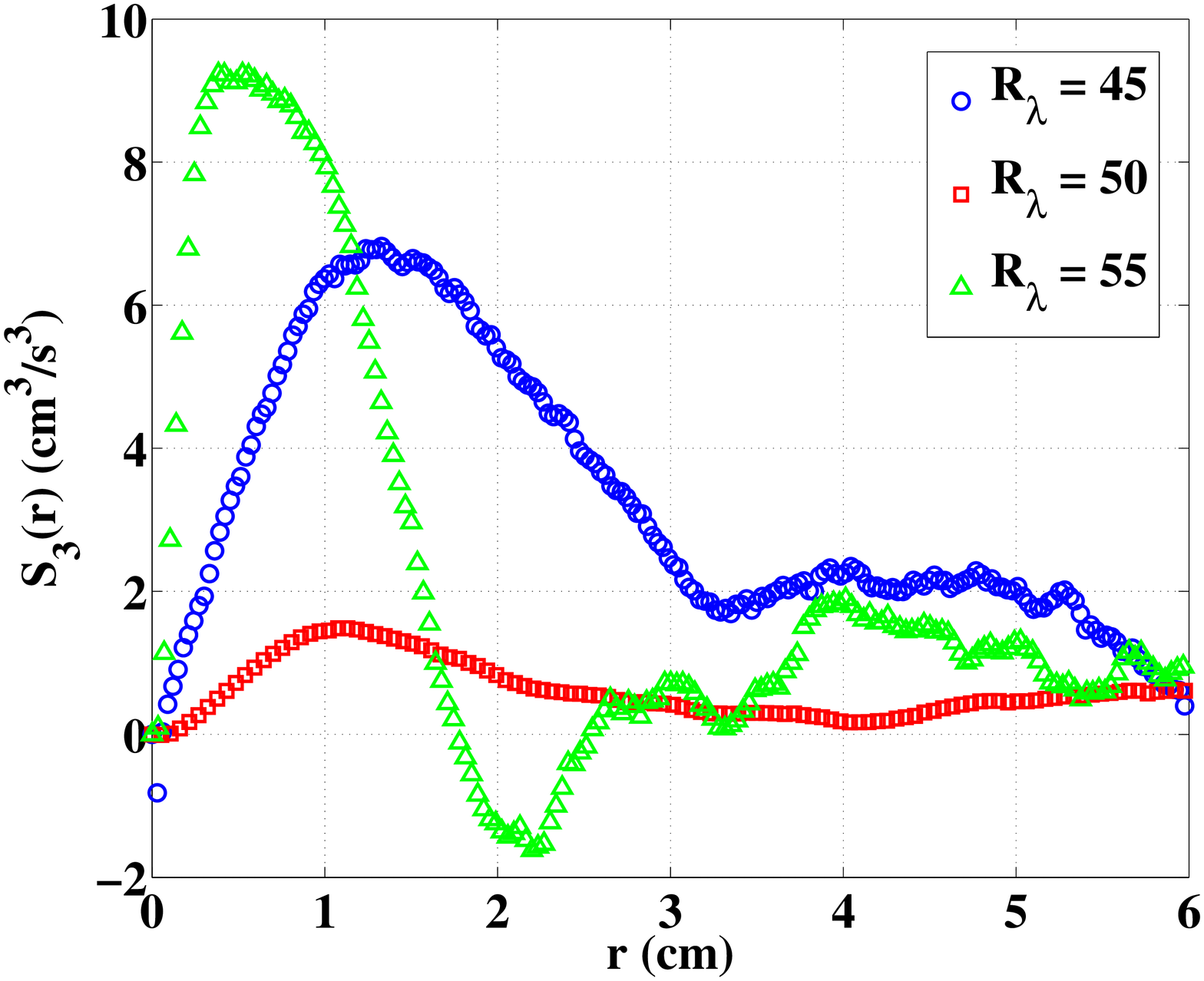}
\caption{The third order structure function $S_3(r)$ vs. $r$ for several $R_{\lambda}$. The sign of $S_3(r)$ is positive for $r$ in the inertial range of each case. This indicates that energy is being transferred to large scales, in agreement with the prediction for the inverse energy cascade of 2D turbulence.}
\label{thirdmoment}
\end{figure}

The sign of third order structure function $S_3(r)$ indicates the direction of energy transfer in the turbulent cascade. (Here the absolute value is not taken.) If $S_3(r) < 0$ for $r$ in the inertial range, then the energy is going to small scales, as in 3D \cite{frisch1995}. If $S_3(r) > 0$, then the energy is going to large scales, as is predicted for the inverse energy cascade of 2D turbulence \cite{kraichnan1980,boffetta2012,kellay2002}. The experimental data that exhibit the energy spectrum scaling $E(k) \propto k^{-5/3}$ also show this positive third moment, as indicated in Fig. \ref{thirdmoment}.

The curves in this figure are positive for $r$ in the inertial range and even appear to behave linearly for small $r$ as the theory suggests. At larger $r$ outside the inertial range they approach zero.

\section{Appendix B: Data Quality}

\begin{figure}[!h]
\centering
\includegraphics[width=0.6\textwidth]{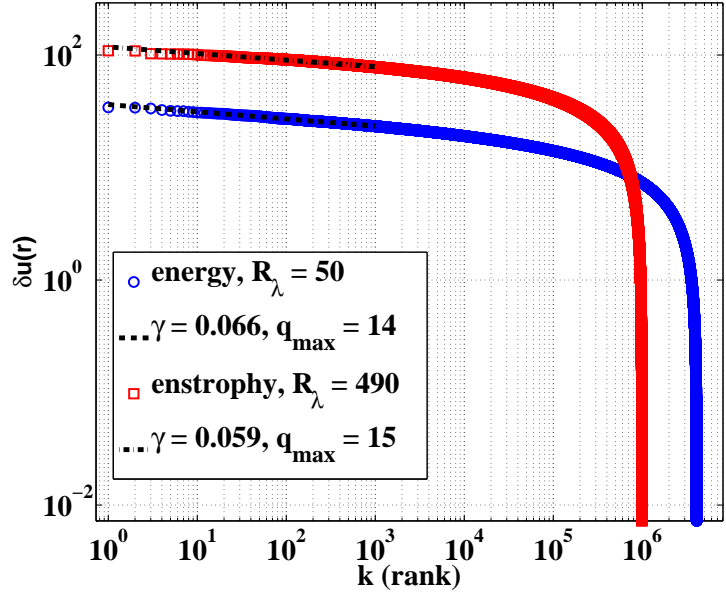}
\caption{Rank-ordered log-log plot of velocity differences for two sets of experimental data. The curves initially behave as power laws. The power law exponent may be used to estimate the highest order of structure functions that can be accurately measured.}
\label{dudok}
\end{figure}

The method of Dudok de Wit for determining the quality of data for measuring structure functions involves an empirical observation \cite{wit2004}. Take all of the measured velocity differences $\delta u(r)$and sort them according to size in descending order $\delta u(r)_k$ ($k$ = 1, 2, 3,...). When plotted in a log-log plot against their ranking, one finds that the distribution is initially a power law $\delta u(r)_k = \alpha {\Big (} \frac{k}{N} {\Big )} ^{-\gamma}$ for roughly $M$ out of the first $N$ elements. This is shown in Fig. \ref{dudok} and has been found to be the general empirical rule for turbulence data.

The structure function can then be broken up according to
\begin{equation}
S_n(r) = \frac{1}{N} \sum_{k=1}^N (\delta u(r)_k)^n = \frac{1}{N} \sum_{k=1}^M (\delta u(r)_k)^n + \frac{1}{N} \sum_{k=M+1}^N (\delta u(r)_k)^n = \frac{1}{N} \sum_{k=1}^M \alpha^n {\Big ( } \frac{k}{N} {\Big )} ^{-n\gamma} + \frac{1}{N} \sum_{k = M+1}^N (\delta u(r)_k)^n 
\end{equation}
The first term is the power law contribution. This represents rare events, since empirically one finds that only a small fraction of the data follow this power law. A simple criterion for the data to have sufficient quality is for this sum to converge as $N$ increases. This leads to the requirement that $n\gamma < 1$. In other words, the highest order moment for which this converges is $n_{max} = \Big{ [} \frac{1}{\gamma} {\Big ]} -1$, where the brackets denote rounding to the nearest integer. We took an average $\gamma$ from all scales $r$ in the inertial range to find $n_{max}$. Using this method, most data were reliable up to order 12 or 13. The main results in this paper only require accuracy up to order 6.

\begin{figure}[h!]
\hfill
\subfigure[$$ n = 1]{\includegraphics[width=5.1cm]{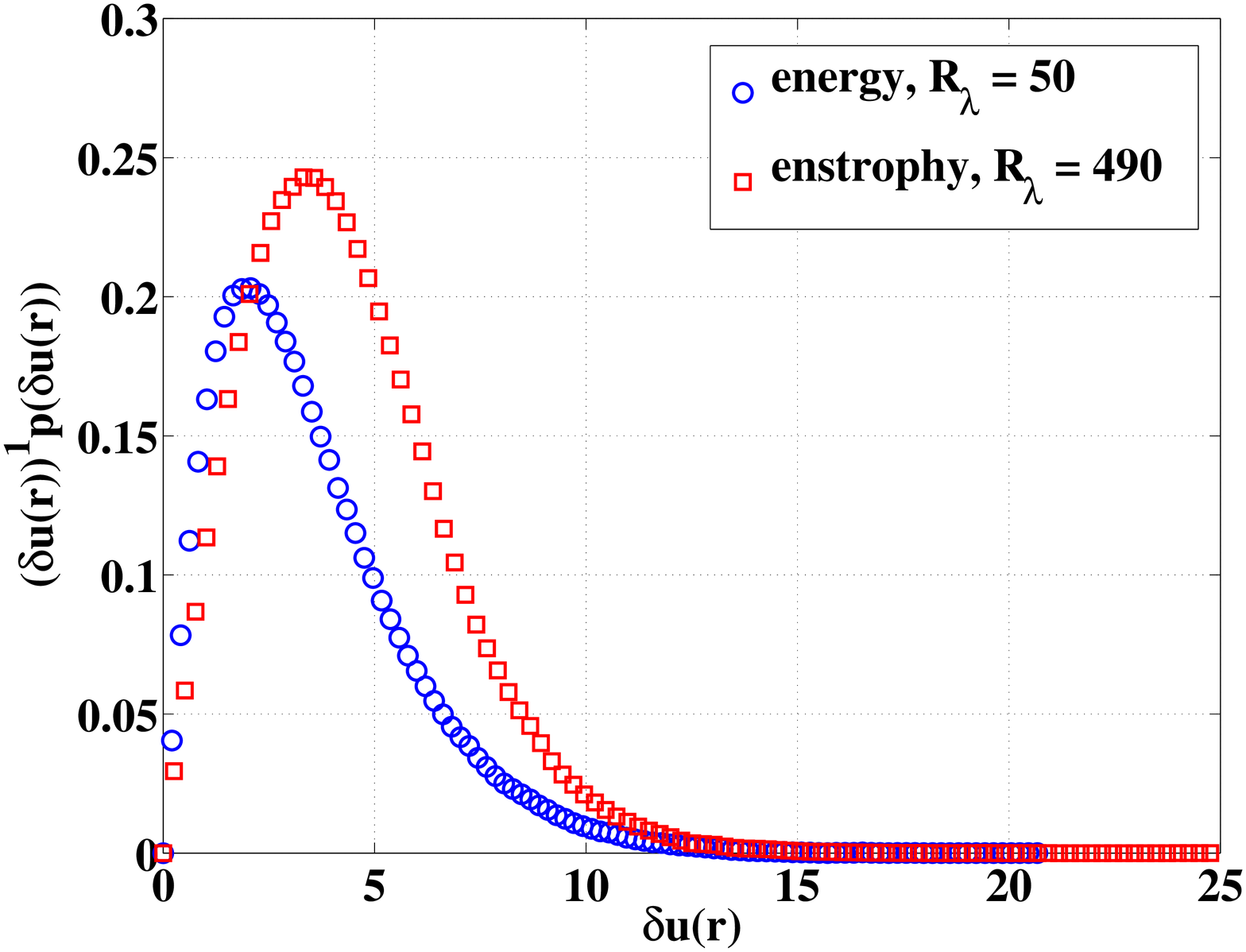}}
\hfill
\subfigure[$$ n = 2]{\includegraphics[width=5.1cm]{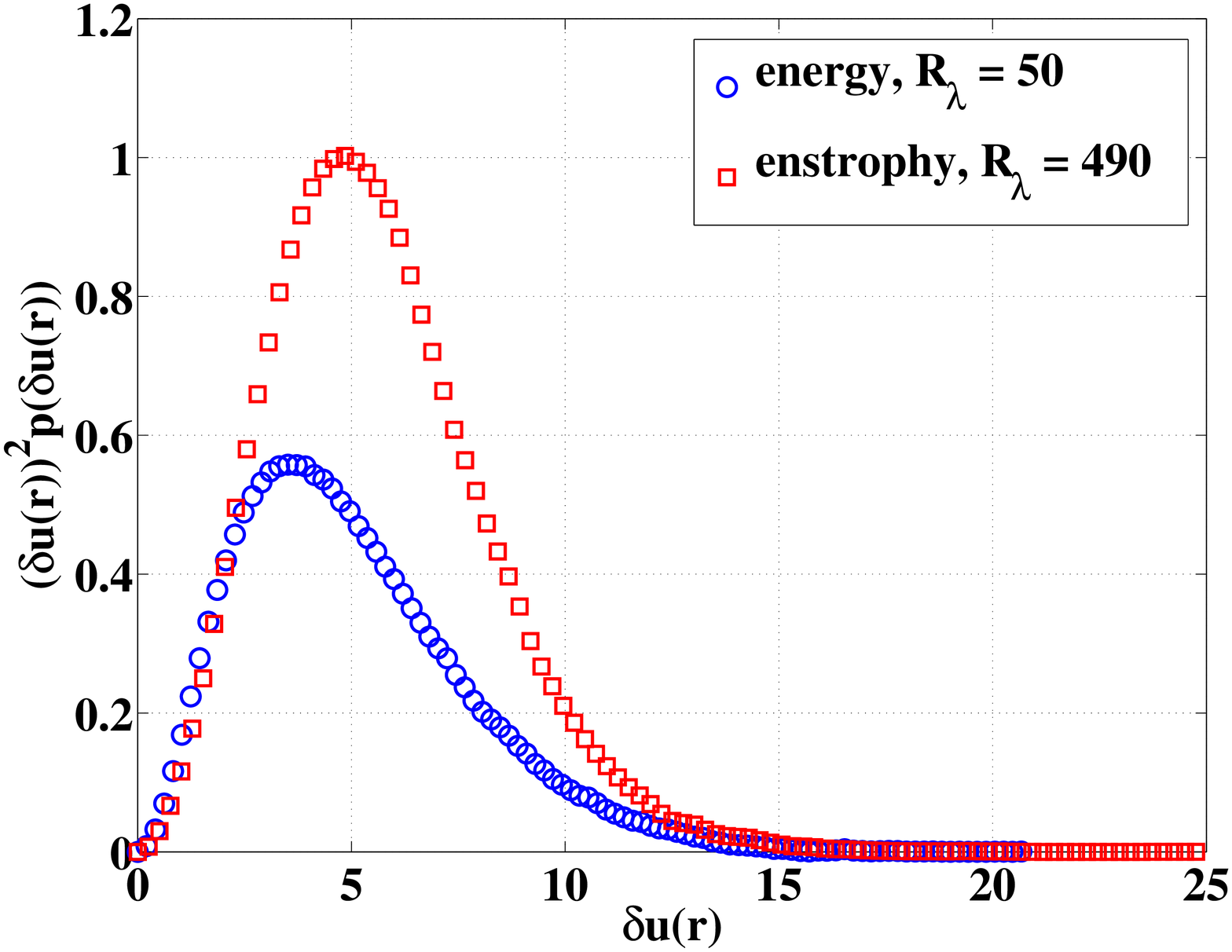}}
\hfill
\subfigure[$$ n = 6]{\includegraphics[width=5.1cm]{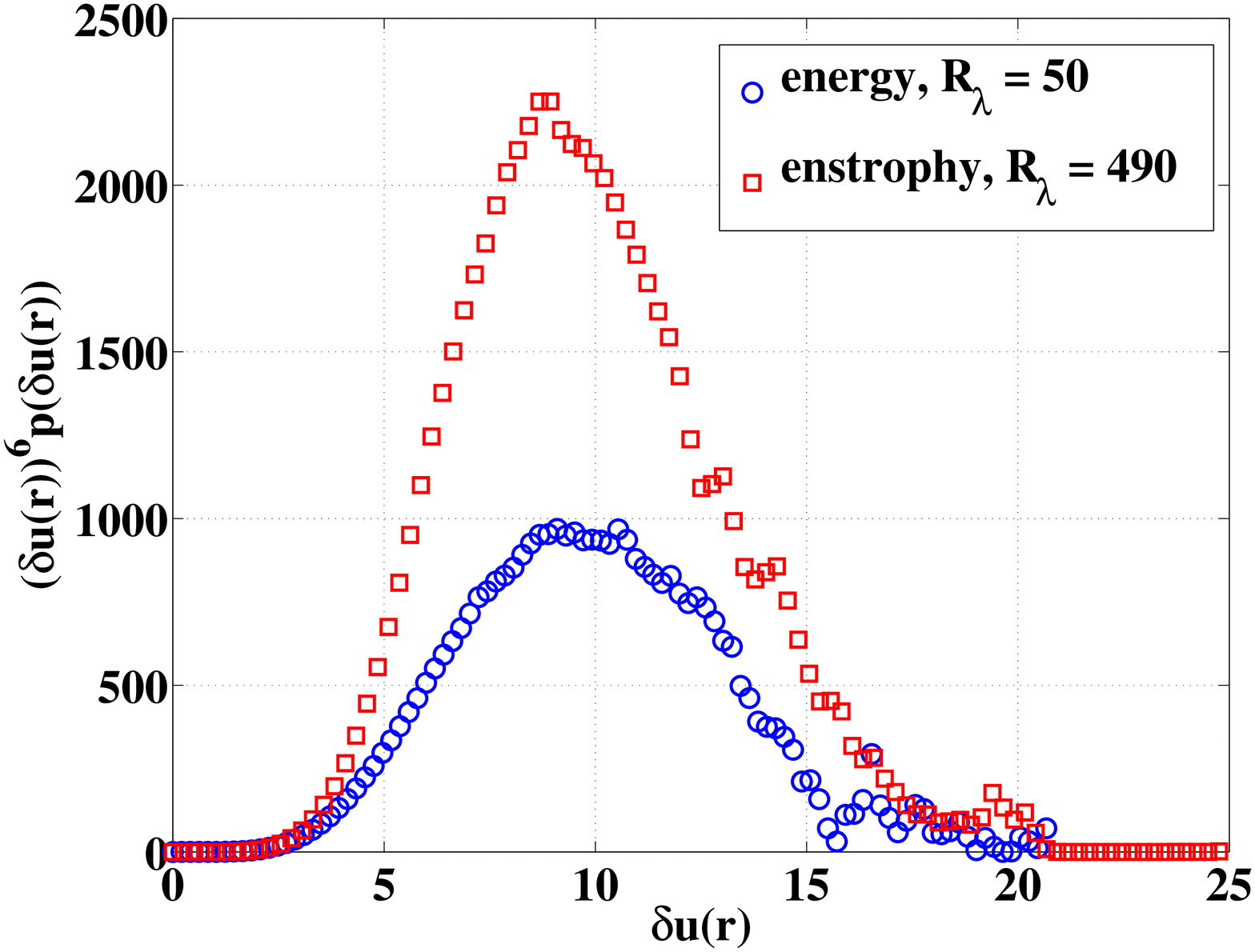}}
\hfill
\caption{The structure function integrand of order $n$ = 1,2 and 6 plotted versus the velocity difference. The area under the curves is finite, indicating that this order structure function can be accurately measured.}
\label{tennekes}
\end{figure}

As a check, we use a second method to test the quality of our data. Since the structure functions are defined as
\begin{equation}
S^n(r) = \int_{-\infty}^{+\infty} \delta u(r)^n p(\delta u(r)) d (\delta u(r)),
\end{equation}
the area underneath the integrand curve $\delta u(r)^n p(\delta u(r))$ must be finite. That is, the integrand plotted as a function of $\delta u(r)$ must not diverge. This is easily checked, although it has the failure of being a less objective method as compared to the method of Dudok de Wit. In Fig. \ref{tennekes} we plot the integrand for several values of $n$ with $r$ chosen as small as possible. The results are the same for larger $r$ values.

The areas underneath the integrand curves are clearly finite. This indicates that the structure functions may be accurately measured up to order 6. Higher orders converged as well, but these plots are not shown here since only order 6 is necessary for the main results of this paper.


\begin{thebibliography}{1}

\bibitem{frisch1995} U. Frisch, {\it Turbulence: The Legacy of A. N. Kolmogorov} (Cambridge University Press, UK 1995)

\bibitem{kraichnan1967a} R. H. Kraichnan, ``Inertial ranges in two-dimensional turbulence," Phys. Fluids {\bf 10}, 1417 (1967)

\bibitem{kraichnan1980} R. H. Kraichnan and D. Montgomery, ``Two-dimensional turbulence," Rep. Prog. Phys. {\bf 43}, 547 (1980)

\bibitem{boffetta2012} G. Boffetta and R. Ecke, ``Two-dimensional turbulence," Ann. Rev. Fluid Mech. {\bf 44}, 427 (2012)

\bibitem{kellay2002} H. Kellay and W. I. Goldburg, ``Two-dimensional turbulence: a review of some recent experiments," Rep. Prog. Phys. {\bf 65}, 845 (2002)

\bibitem{tabeling2002a} P. Tabeling, ``Two-dimensional turbulence: a physicist approach," Phys. Rep. {\bf 362}, 1 (2002)

\bibitem{chen2006} S. Chen, R. E. Ecke, G. L. Eyink, M. Rivera, M. Wan, and Z. Xiao, ``Physical mechanism of the two-dimensional inverse energy cascade," Phys. Rev. Lett. {\bf 96}, 084502 (2006)

\bibitem{chen2003} S. Chen, R. E. Ecke, G. L. Eyink, X. Wang, and Z. Xiao, ``Physical mechanism of the two-dimensional enstrophy cascade," Phys. Rev. Lett. {\bf 91}, 214501 (2003)

\bibitem{boffetta2007} G. Boffetta, ``Energy and enstrophy fluxes in the double cascade of two-dimensional turbulence," J. Fluid Mech. {\bf 589}, 253 (2007)

\bibitem{boffetta2010} G. Boffetta and S. Musacchio, ``Evidence for the double cascade scenario in two-dimensional turbulence," Phys. Rev. E {\bf 82}, 016307 (2010)

\bibitem{rutgers1998} M. A. Rutgers, ``Forced 2D turbulence: experimental evidence of simultaneous inverse energy and forward enstrophy cascades," Phys. Rev. Lett. {\bf 81}, 2244 (1998)

\bibitem{daniel2000} W. B. Daniel and M. A. Rutgers, ``Intermittency in forced two-dimensional turbulence,"  arXiv:nlin/0005008

\bibitem{bruneau2005} C. H. Bruneau and H. Kellay, ``Experiments and direct numerical simulations of two-dimensional turbulence," Phys. Rev E {\bf 71}, 046305 (2005)

\bibitem{kolmogorov1941} A. N. Kolmogorov, ``The local structure of turbulence in incompressible viscous fluids for very large Reynolds numbers," Dokl. Akad. Nauk. SSSR {\bf 30}, 299 (1941) (Proc. R. Soc. Lond. A {\bf 434} (reprinted))

\bibitem{kraichnan1967b} R. H. Kraichnan, ``Intermittency in the very small scales of turbulence,"Phys. Fluids {\bf 10}, 2080 (1967)

\bibitem{sreenivasan1997} K. R. Sreenivasan and R. A. Antonia, ``The phenomenology of small-scale turbulence," Ann Rev. Fluid Mech. {\bf 29}, 435 (1997)

\bibitem{boffetta2000} G. Boffetta, A. Celani, and M. Vergassola, ``Inverse energy cascade in two-dimensional turbulence: deviations from gaussian behavior," Phys. Rev. E {\bf 61}, R29 (2000)

\bibitem{paret1998} J. Paret and P. Tabeling, ``Intermittency in the two-dimensional inverse cascade of energy: experimental observations," Phys. Fluids {\bf 10}, 3126 (1998)

\bibitem{jun2005} Y. Jun and X. L. Wu, ``Large-scale intermittency in two-dimensional driven turbulence," Phys. Rev. E {\bf 72}, 035302 (2005)

\bibitem{boffetta2002} G. Boffetta, A. Celani, S. Musacchio, and M. Vergassola, ``Intermittency in two-dimensional Ekman-Navier-Stokes turbulence," Phys. Rev. E {\bf 66}, 026304 (2002)

\bibitem{nam2000} K. Nam, E. Ott, T. M. Antonsen Jr., and P. N. Guzdar, ``Lagrangian chaos and the effect of drag on the enstrophy cascade in two-dimensional turbulence," Phys. Rev. Lett. {\bf 84}, 5134 (2000)

\bibitem{tsang2005} Y. K. Tsang, E. Ott, T. M. Antonsen Jr., and P. N. Guzdar, ``Intermittency in two-dimensional turbulence with drag," Phys. Rev. E {\bf 71}, 066313 (2005)

\bibitem{paret1999} J. Paret, M. C. Jullien, and P. Tabeling, ``Vorticity statistics in the two-dimensional enstrophy cascade," Phys. Rev. Lett. {\bf 83}, 3418 (1999)

\bibitem{boffetta2005} G. Boffetta, A. Cenedese, S. Espa, and S. Musacchio, ``Effects of friction on 2D turbulence: an experimental study of the direct cascade," Europhys. Lett. {\bf 71}, 590 (2005)

\bibitem{bracco2010} A. Bracco and J. C. McWilliams, ``Reynolds-number dependency in homogeneous, stationary two-dimensional turbulence," J. Fluid Mech. {\bf 646}, 517 (2010)

\bibitem{benzi1986} R. Benzi, G. Paladin, S. Patarnello, P. Santangelo, and A. Vulpiani, ``Intermittency and coherent structures in two-dimensional turbulence," J. Phys. A: Math. Gen. {\bf 19}, 3771 (1986)

\bibitem{jimenez2007} J. Jimenez, ``Intermittency in turbulence," Proc. 15th 'Aha Huliko' Winter Workshop, Extreme Events,  81,  (2007)

\bibitem{chakraborty2011} P. Chakraborty, T. Tran, and G. Gioia, ``Leading-order dynamics of gravity-driven flows in free-standing soap films," J. Elasticity {\bf 104}, 105 (2011)

\bibitem{kahalerras1998} H. Kahalerras, Y. Mal\'{e}cot, Y. Gagne, and B. Castaing, ``Intermittency and Reynolds number," Phys. Fluids {\bf 10}, 910 (1998)

\bibitem{tran2010} T. Tran, P. Chakraborty, N. Guttenberg, A. Prescott, H. Kellay, W. I. Goldburg, N. Goldenfeld, and G. Gioia, ``Macroscopic effects of the spectral structure in turbulent flows," Nat. Phys. {\bf 6}, 438 (2010)

\bibitem{merzkirch1987} W. Merzkirch, {\it Flow Visualization} (Academic, Orlando, 1987)

\bibitem{belmonte2000} A. Belmonte, B. Martin, and W. I. Goldburg, ``Experimental study of Taylor’s hypothesis in a turbulent soap film," Phys. Fluids {\bf 12}, 835 (2000)

\bibitem{ramond2000} A. Ramond and P. Millan, ``Measurements and treatment of LDA signals, comparison with hot-wire signals," Expts. Fl. {\bf 28}, 58 (2000)

\bibitem{pope2000} S. A. Pope, {\it Turbulent Flows} (Cambridge University Press, Cambridge, 2000)

\bibitem{kellay2012} H. Kellay, T. Tran, W. I  Goldburg, N. Goldenfeld, G. Gioia, and P. Chakraborty, ``Testing a missing spectral link in turbulence," Phys. Rev. Lett. {\bf 109}, 254502 (2012)

\bibitem{belmonte1999} A. Belmonte, W. I. Goldburg, H. Kellay, M. A. Rutgers, B. Martin, and X. L. Wu, ``Velocity fluctuations in a turbulent soap film: The third moment in two dimensions," Phys. Fluids {\bf 11}, 1196 (1999)

\bibitem{tabeling2002b} P. Tabeling and H. Willaime, ``Transition at dissipative scales in large-Reynolds-number turbulence," Phys. Rev. E {\bf 65}, 066301 (2002)

\bibitem{tran2006} C. V. Tran and D. G. Dritschel, ``Vanishing enstrophy dissipation in two-dimensional Navier–Stokes turbulence in the inviscid limit," J. Fluid Mech. {\bf 559}, 107 (2006)

\bibitem{dritschel2007} D. G. Dritschel, C. V. Tran, and R. K. Scott, ``Revisiting Batchelor's theory of two-dimensional turbulence," J. Fluid Mech. {\bf 591}, 379 (2007)

\bibitem{wit2004} T. Dudok de Wit, ``Can high-order moments be meaningfully estimated from experimental turbulence measurements?," Phys. Rev.  E {\bf 70}, 055302 (1993)

\bibitem{benzi1993} R. Benzi, R. Tripiccione, C. Baudet, F. Massaioli, and S. Succi, ``Extended self-similarity in turbulent flows," Phys. Rev.  E {\bf 48}, R29 (1993)

\bibitem{anselmet1984} F. Anselmet, Y. Gagne, E. J. Hopfinger, and R. J. Antonia, ``High-order velocity structure functions in turbulent shear flows,"  J. Fluid Mech. {\bf 140}, 63 (1984)

\bibitem{rivera2000} M. Rivera and X. L. Wu, ``External dissipation in driven two-dimensional turbulence," Phys. Rev. Lett. {\bf 85}, 976 (2000)

\bibitem{martin1998} B. K. Martin, X. L. Wu, W. I. Goldburg and M. A. Rutgers, ``Spectra of decaying turbulence in soap films," Phys. Rev. Lett. {\bf 80}, 3964 (1998)

\end{thebibliography}
\end{document}